# Multitemporal Latent Dynamical Framework for Hyperspectral Images Unmixing

Ruiying Li, Bin Pan, Lan Ma, Xia Xu and Zhenwei Shi

*Abstract*—Multitemporal hyperspectral unmixing can capture dynamical evolution of materials. Despite its capability, current methods emphasize variability of endmembers while neglecting dynamics of abundances, which motivates our adoption of neural ordinary differential equations to model abundances temporally. However, this motivation is hindered by two challenges: the inherent complexity in defining, modeling and solving prob– lem, and the absence of theoretical support. To address above challenges, in this paper, we propose a multitemporal latent dynamical (MiLD) unmixing framework by capturing dynamical evolution of materials with theoretical validation. For address– ing multitemporal hyperspectral unmixing, MiLD consists of problem definition, mathematical modeling, solution algorithm and theoretical support. We formulate multitemporal unmixing problem definition by conducting ordinary differential equations and developing latent variables. We transfer multitemporal unmixing to mathematical model by dynamical discretization approaches, which describe the discreteness of observed sequence images with mathematical expansions. We propose algorithm to solve problem and capture dynamics of materials, which approximates abundance evolution by neural networks. Fur– thermore, we provide theoretical support by validating the crucial properties, which verifies consistency, convergence and stability theorems. The major contributions of MiLD include defining problem by ordinary differential equations, modeling problem by dynamical discretization approach, solving problem by multitemporal unmixing algorithm, and presenting theoretical support. Our experiments on both synthetic and real datasets have validated the utility of our work.

## I. Introduction

HYPERSPECTRAL image contains significant spectral information, leading to various hyperspectral data processing tasks for extracting valuable knowledge [1]–[6]. However, low spatial resolution existing in hyperspectral image causes mixing pixels, which hinders the identification of materials [7]–[10]. Hyperspectral unmixing can solve mixing pixels problem by extracting pure spectrum (*Endmembers*) and their corresponding coefficients (*Abundances*) [11], [12]. Methods of unmixing generally depend on different mixing models that can be categorized into linear or nonlinear types [13]–[16]. In which linear mixing model (LMM) describes the observed pixel as a weighted sum of endmembers, with the weights corresponding to abundances [17].

Due to the simplicity and interpretability, LMM is classically adopted for addressing unmixing problem [18]–[22]. Methods based on LMM can be characterized as several types, including geometrical, statistical, nonnegative matrix factorization, sparse regression, and deep learning methods [23]. Most of the above methods focus on a single observed hyperspectral image, which is referred to as single-phase hyperspectral image unmixing. Despite its potential, single-phase unmixing is limited in providing comprehensive insights into the temporal evolution of materials.

Recently, multitemporal hyperspectral unmixing (MTHU) has attracted considerable interest due to its potential in analyzing temporal information. Given the ability of sensors to acquire time-series hyperspectral images across multiple temporal domains, MTHU intends to develop methods for capturing the evolution of endmembers with their abundances [24], [25]. As a result of its dynamics capabilities, multitemporal unmixing is widely used in many applications [26]–[28].

Current studies on multitemporal unmixing primarily emphasize the exploration of endmembers, which is tightly related to, e.g., spectral variability [24], [29] and dynamics of the endmembers [30], [31]. In addition, models derived from a statistical perspective have been developed. For instance, Bayesian methods are adopted in multitemporal unmixing to incorporate composite prior information while infer dynamical evolution of materials [31]–[33]. In particular, Wang et al. extends spectral unmixing to the spatio-temporal domain [34]. A transformer structure for multitemporal unmixing is proposed to capture rich multi-dimensional information [35]. Above methods demonstrated their proficiency in modeling the dynamical evolution of endmembers.

Despite the promising results achieved by previous studies, they may potentially overlook the dynamical evolution of abundances. Current researches typically model abundances either as simply linear transformation across temporal (e.g., $A_t = A_{t-1} + \Delta_t$) or by specifying a conditional probability (e.g., $A_t \sim P_\theta(A_t | A_{t-1})$). The former may be insufficient to capture time-dependent variation of abundances, while the latter presents challenges in specification and incurs higher computational complexity. These limitations may create obstacles for abundances estimation, potentially leading to inaccuracies in the interpretation of material dynamics over time.

The work was supported by the National Key Research and Development Program of China under Grant 2022YFA1003800, the National Natural Science Foundation of China under the Grant 62125102, and the Fundamental Research Funds for the Central Universities under grant 63243074. (Corresponding author: Bin Pan)

Ruiying Li, Bin Pan (corresponding author) and Lan Ma are with the School of Statistics and Data Science, KLMDASR, LEBPS, and LPMC, Nankai University, Tianjin 300071, China. (e-mail: liruiying@mail.nankai.edu.cn; panbin@nankai.edu.cn; malan@mail.nankai.edu.cn).

Xia Xu is with the School of Computer Science and Technology, Tiangong University, Tianjin 300387, China (e-mail: xuxia@nankai.edu.cn)

Zhenwei Shi is with the Image Processing Center, School of Astronautics, and the State Key Laboratory of Virtual Reality Technology and Systems, Beihang University, Beijing 100191, China (e-mail: shizhenwei@ buaa.edu.cn).

Consequently, more effective results should be achieved by considering abundances temporally.

The motivations of our method are modeling abundances and endmembers in a dynamical manner jointly. In order to directly operate abundances, we represent abundances as latent variables by considering autoencoder, a widely used tool for unmixing [36]–[42]. Moreover, inspired by neural ordinary differential equations (neural ODEs) [43]–[46], which provide powerful algorithms to learn the temporal dynamics for time series modeling (see Section II for a review). This paper aims at developing a latent dynamical framework building upon neural ODEs to describe the evolution of abundances and model the variability of endmembers.

However, utilizing neural ordinary differential equation for multitemporal unmixing may suffer two challenges:

1) How to construct effective problem definition, mathematical model, and solution algorithm. The discreteness of images is opposite to the continuity of ODEs. Current simply ODEs models limit their ability to capture complex dynamics.
2) How to provide theoretical support. The important properties of ODEs have not been guaranteed in addressing multitemporal hyperspectral unmixing problem.

In order to address above issues, in this paper, we propose a **Mi**ltitemporal **L**atent **D**ynamical (MiLD) unmixing framework by capturing dynamics of materials with theoretical validation. MiLD defines, models, solves and supports multitemporal hyperspectral unmixing problem. To define problem, we represent abundances as latent variables and conduct ordinary differential equations whose differential formulation is redesigned by two properties. To model problem, we develop dynamical discretization approaches which describe discreteness of irregular images by mathematical expansions and improvements. To solve problem, we propose multitemporal unmixing algorithm which approximates complex dynamical evolution of materials by neural networks. Moreover, to provide theoretical support, we demonstrate consistency and convergence theorems for modeling, and stability theorem for algorithm. The main contributions of MiLD can be summarized as follows:

1) We propose problem formulations to define multitemporal unmixing, which conduct ordinary differential equations and represent abundance as latent variables.
2) We propose dynamical discretization approach for problem modeling, which describe discreteness of multitemporal images by mathematical expansions.
3) We propose an algorithm for multitemporal unmixing, which solves problem and captures abundances dynamics with endmembers variabilities.
4) We provide theoretical support, which validates the crucial consistency, convergence and stability properties for our framework.

The remainder of this paper is structured as follows. Section II describes the related work. Section III introduces the proposed algorithm and validate properties. Section V describes the experimental results on both synthetic datasets and real datasets.

## II. Related Work

### A. Multitemporal Hyperspectral Unmixing

Multitemporal hyperspectral image analysis is crucial for investigating material variations temporally. However, tradeoff between spatial and spectral resolution in sequence hyperspectral images complicates the accurate pixel identification, making multitemporal unmixing increasingly important.

Let $\{\mathbf{Y_t}\}_{t=1}^{T}$ denote multitemporal hyperspectral images sequence including $T$ time, where $t$ is the discrete time index and $\mathbf{Y_t}$ is the observed hyperspectral image at time $t$. If the dimensions of $\mathbf{Y_t}$ can be described as $\mathbf{Y_t} \in R^{H \times W \times L}$, thus $\{\mathbf{Y_t}\}_{t=1}^{T} \in R^{T \times H \times W \times L}$. $H$, $W$ and $L$ represents height, weight and number of channels, respectively. At each time $t$, we consider the classical linear mixing model and model $\mathbf{Y_t}$ as the following equation:

$$\mathbf{Y_t} = \mathbf{A_t} \mathbf{E_t} + \mathbf{N_t} \qquad (1)$$

where $\mathbf{A_t} \in R^{H \times W \times P}$ and $\mathbf{E_t} \in R^{P \times L}$ denote abundance and endmember at time $t$. $\mathbf{A_t}$ obeys the popular nonnegative constraint and sum-to-one constraint. $P$ is the number of spectral signatures, which is assumed known previously.

The observed sequence of images $\{\mathbf{Y_t}\}_{t=1}^{T}$ is extracted from the same spatial scenes at different times. Thus these images are likely to share similarities. Such correlation can be effectively captured by modeling the dynamics of materials, which contains endmember variability and dynamics of abundance. The two aspects can be simply described as:

$$\hat{\mathbf{E}}_\mathbf{t} = f_E(\hat{\mathbf{E}}_{\mathbf{t-1}}) \qquad (2)$$
$$\hat{\mathbf{A}}_\mathbf{t} = f_A(\hat{\mathbf{A}}_{\mathbf{t-1}}) \qquad (3)$$

$f_E$ and $f_A$ express the evolution of materials from time $t-1$ to time $t$, which vary across different studies.

*1) Endmember Variability:* As materials change over time, endmembers exhibit dynamical evolution temporally, which can be approximated by endmember variability at time $t$ [47]–[52]. Among them, endmember variability can be modeled as an additive perturbation, following perturbed linear mixing model (PLMM) [32], [48], that is

$$\hat{\mathbf{E}}_\mathbf{t} = \mathbf{E} + d\hat{\mathbf{E}}_\mathbf{t} \qquad (4)$$

where $\mathbf{E} \in R^{P \times L}$ denotes the reference endmembers, $d\hat{\mathbf{E}}_\mathbf{t} \in R^{P \times L}$ are the perturbation vectors associated with the $t$th image. Then $\mathbf{Y_t}$ can be expressed as:

$$\mathbf{Y_t} = \mathbf{A_t}(\mathbf{E} + d\hat{\mathbf{E}}_\mathbf{t}) + \mathbf{N}_t \qquad (5)$$

*2) Dynamics of Abundance:* Most existing algorithms focus primarily on endmember variability while simplifying the changes in abundance. Methods mentioned for abundance dynamics can be divided into two categories: simple linear transformations and statistical methods. The former operation can be formulated as [29]:

$$\mathbf{A_t} = \mathbf{A_{t-1}} + \Delta_t \qquad (6)$$

if $\Delta_t = 0$, it satisfies that abundances are constant in time [30]. While the statistical methods provides conditional probability or prior distributions [31]–[33]. Some of them can be expressed as:

$$A_t \sim P_\theta(A_t|A_{t-1}) \tag{7}$$

where $P_\theta$ denotes potential distribution function and mostly is set as Gaussian distribution.

However, the simply linear transformation may not enough to model the dynamical evolution of abundances. Moreover, conditional probability information is hard to specify and usually accompanies with higher computational complexity. Consequently, methods that emphasize both endmember and abundance must be investigated.

### B. Neural Ordinary Differential Equations

Neural Ordinary Differential Equations (Neural ODEs) are a family of continuously-defined latent time dynamics, which provides novel algorithms for modeling temporal dynamics [43]. These models define hidden state $h(t)$ as the solution to an ordinary differential equation and parameterize the derivative of $h(t)$ by developing a neural network, thus $h(t)$ is defined as:

$$\frac{dh(t)}{dt} = f_\theta(h(t), t) \tag{8}$$

where $f$ is a neural network with parameters $\theta$ that defines the dynamical change of an temporal system. The framing of $h(t)$ is a key module, which helps to capture time dependence and dynamic properties of sequences. Among existing methods, ODE-RNNs and Latent ODEs can naturally handle arbitrary time gaps between observations, and can explicitly model the probability of observation times [44]. Neural controlled differential equation model is directly applicable to the general setting of partially observed irregularly-sampled multivariate time series [45]. In neural latent dynamics model [46], the latent representations evolving with an ODE is enhanced by the change of observed signal $\{x_t\}_{t=0}^T$:

$$\frac{dh(t)}{dt} = f_\theta(h(t-1), \{x_k\}_{k=0}^{t-1}, t) \tag{9}$$

The powerful ability of Neural ODEs to model latent dynamics makes them a promising approach for temporal data analysis. Building on this capability, we introduce ODEs as an algorithm for capturing dynamiccal evolution of materials in multi-temporal hyperspectral unmixing. However, existing ODEs architectures for multi-temporal unmixing are limited. The continuity of ODEs is conflict with discrete hyperspectral images. Moreover, consistency, convergence and stability are not guaranteed.

## III. METHODOLOGY

In this section, we introduce the proposed MiLD in detail. In Section III-A, we define multitemporal hyperspectral unmixing by ordinary differential equations (see Eq.(11)) in latent space. In Section III-B, we describe multitemporal hyperspectral unmixing as mathematical model and discretize Eq.(11) to Eq.(21). In Section III-C, we make improvements. In Section III-D, we design algorithm to solve problem. In Section III-E, we demonstrate theoretical support. The flowchart of MiLD is presented in Fig.(1).

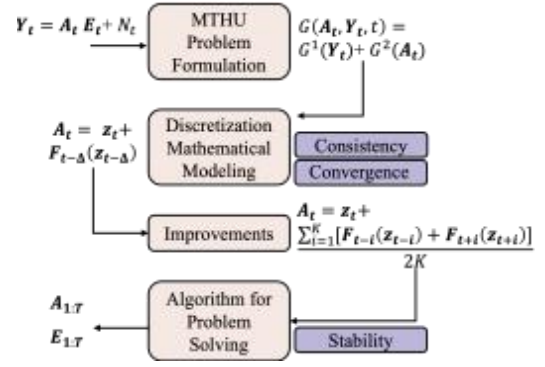

Fig. 1. Flowchart of the proposed MiLD framework. MiLD consists of problem definition, mathematical modeling, solution algorithm and theoretical support. Consistency and convergence theorems support mathematical modeling process. Stability theorem supports the algorithm.

### A. Latent Variables–based Problem Formulation for MTHU

In this subsection, we give problem definition for multitemporal hyperspectral unmixing by adopting ordinary differential equations. Abundances are regarded as latent variables by conducting autoencoders. Moreover, we expand the dynamics of abundance and incorporate the change of observations to learn more complex evolution.

We first consider multitemporal unmixing problem with observed sequence images $\{Y_t\}_{t=0}^T$ as multiple single unmixing tasks (see Fig.2(b)). By developing $T$ auto-encoders independently, $T$ single unmixing tasks can be described as:

$$\{\approx_t\}_{t=1}^T = \text{Encoder}_{1:T}\left(\{Y_t\}_{t=1}^T\right) \tag{10}$$
$$\text{s.t. } Y_t = \approx_t E_t$$

where $\{\approx_t\}_{t=1}^T \in \mathbb{R}^{T \times H \times W \times P}$ are pseudo-abundances across $T$ moments and can be viewed as latent variables. However, $\approx_t$ is estimated independently, which ignores the correlation between each time and the dynamics of abundance. Thus, utilizing $\approx_t$ may cause potential inaccuracies in multitemporal unmixing problem.

To captures dynamical evolution of materials, we improve $\approx_t$ to get real abundances $A_t$ in latent space. Under this operation, pseudo-abundance $\approx_t$ can be viewed as an approxi- mation of real abundance $A_t$ without feature fusion from other phrases.

As neural ordinary differential equations are attractive for modeling temporal dynamics, this paper aims to construct a latent dynamical system for describing dynamics of $A_t$. We describe it as neural ordinary differential equation:

$$\frac{dA_t}{dt} = G(A_t, Y_t, t) \tag{11}$$

As observed hyperspectral image $Y_t$ are related with other phrase $Y_{1:T}$. $\approx_t$ can be related temporally through $Y_t$:

$$\frac{dz_t}{dt} = G^1(Y_t) \tag{12}$$

Our model choice over the latent dynamics $G(\cdot)$ is by two properties: (i) $G(\cdot)$ can expand the latent abundance $A_t$ and (ii) $G(\cdot)$ can describe how the change of hyperspectral images

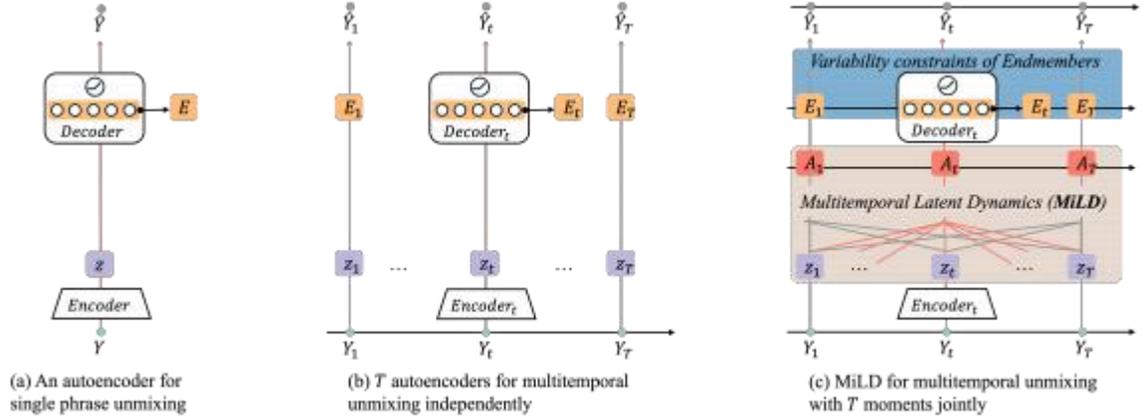

Fig. 2. The sketch of single phrase unmixing, independent multitemporal unmixing and our proposed MiLD for multitemporal unmixing jointly. (a) a simply autoencoder architecture for single phrase unmixing: abundance is latent variable and endmember is viewed as weight of decoder; (b) T autoencoders for multitemporal unmixing independently: addressing sequence images as irrelevant which obtains the T pseudo-abundances; (c) MiLD: taking T multitemporal images jointly to capture dynamical evolution of materials.

influence latent variables. Following the above properties, we assume $G(\cdot)$ is consist of two parts:

$$G(A_t, Y_t, t) = G^1(Y_t) + G^2(A_t) \quad (13)$$

where $G^1(Y_t)$ is mentioned in eq.(12). The change of input images can directly influence the local dynamics of the hidden states. $G^2(A_t)$ specifies the dynamics of $A_t$ in a differentiable manner. Thus, MTHU is formulated as new problem definition through Eq.(11). Solving Eq.(11) is equivalent to addressing MTHU problem.

### B. Dynamical Discretization–based Mathematical Modeling

In this subsection, we model multitemporal hyperspectral unmixing by mathematical expansions and approximations. As the continuity of function $G(\cdot)$ is opposite to the inherently discreteness of hyperspectral images $Y_t$, it is essential to de - velop an appropriate discretization approach. Thus we design new dynamical discretization approach to discretize Eq.(11).

For a desired sequence of discrete hyperspectral images $Y_t$, the preliminary approximations to $A_t$ can be obtained following Trapezoidal rule by iterating the equation:

$$\int_a^b f(x)dx \approx \sum_{k=1}^{N} \frac{f(x_{k-1}) + f(x_k)}{2} \Delta x_k \quad (14)$$

Given the step size $\Delta$, along the traditional left-to-right direction, eq.(11) can be discretized following eq.(14):

$$\begin{aligned} A_t &= A_{t-\Delta} + \frac{\Delta}{2}[G(A_t, Y_t, t) + G(A_{t-\Delta}, Y_{t-\Delta}, t-\Delta)] \\ &= A_{t-\Delta} + \frac{\Delta}{2}[G^2(A_t) + G^2(A_{t-\Delta})] \\ &\quad + \frac{\Delta}{2}[G^1(Y) + G^1(Y_{t-\Delta})] \end{aligned} \quad (15)$$

Following the same operation, eq.(12) can be discretized to:

$$z_t = z_{t-\Delta} + \frac{\Delta}{2}[G^1(Y_t) + G^1(Y_{t-\Delta})] \quad (16)$$

Put this equation into eq.(15):

$$A_t = A_{t-\Delta} + \frac{\Delta}{2}[G^2(A_t) + G^2(A_{t-\Delta})] \quad (17)$$
$$\quad + z_t - z_{t-\Delta}$$

Take Taylor expansion for $G^2(\cdot)$ at time $t - \Delta$:

$$G^2(A_t) = G^2(A_{t-\Delta}) + \Delta \frac{d(G^2(A_{t-\Delta}))}{d(t-\Delta)} \quad (18)$$

Thus eq.(17) equals to:

$$A_t - z_t = A_{t-\Delta} - z_{t-\Delta} \\ + \frac{\Delta}{2}\left[2 \times G^2(A_{t-\Delta}) + \Delta \frac{d(G^2(A_{t-\Delta}))}{d(t-\Delta)}\right] \quad (19)$$

this equation is said to be consistent with the differential equation eq.(11). This consistency will be proved in Section III-D.

For simplifying this equation, we consider two conditions. (i) As mentioned in Section III-A, $z_t$ can be viewed as an approximation of $A_t$ without feature fusion from other phrases. (ii) Sequence $\{z_t\}$ is convergent to sequence $\{A_t\}$ at a small $\Delta$. This convergence will be proved in Section III-D. Based on these conditions, when calculating $A_t$, we approximate other phrase $A_{(1:T)\backslash t}$ by $z_{(1:T)\backslash t}$. Under the approximation, eq.(19) equals to:

$$A_t - z_t = \frac{\Delta}{2}\left[2 \times G^2(z_{t-\Delta}) + \Delta \frac{d(G^2(z_{t-\Delta}))}{d(t-\Delta)}\right] \quad (20)$$

Thus $A_t$ from left-to-right discretization is expressed as:

$$A_t = F_{t-\Delta}(z_{t-\Delta}) + z_t$$
$$where \; F_t(z_t) = \frac{\Delta}{2}\left[2 \times G^2(z_t) + \Delta \frac{d(G^2(z_t))}{d(t)}\right] \quad (21)$$

From this formulation, we could infer that $A_t$ at time t is related with information from other phrase $t - \Delta$ and input hyperspectral image at time t. The first part $F_{t-\Delta}(z_{t-\Delta})$ improves the connection between time t and time $t - \Delta$. The second part $z_t = \text{Encoder}_t(Y_t)$ describes how the change of observed images influence latent variables.

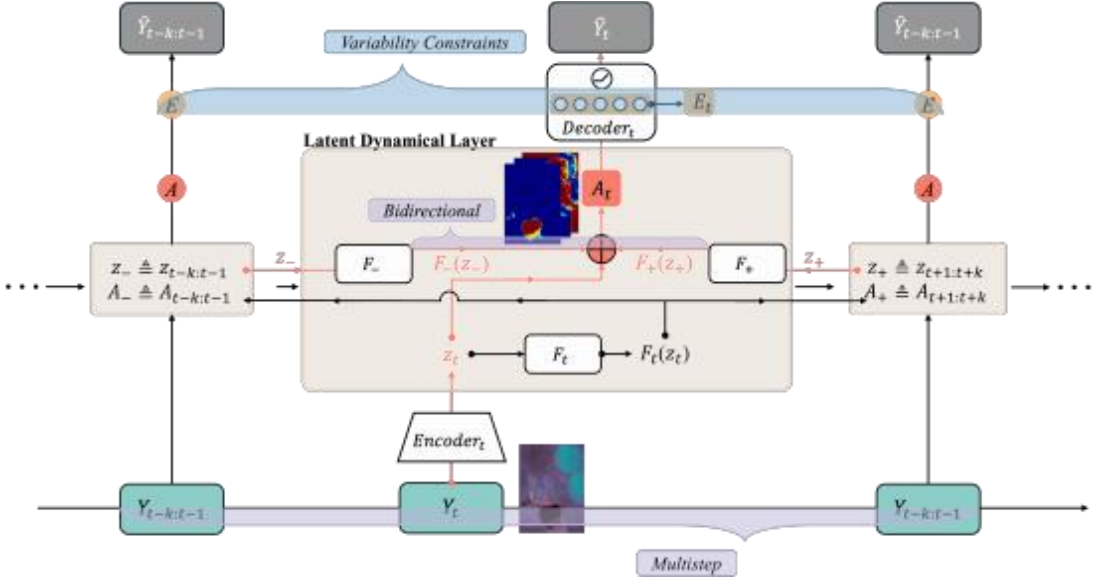

Fig. 3. Overview of the proposed algorithm. From bottom to top are three parts: the calculation of pseudo-abundances $z_t$, the latent dynamical layer for abundance $A_t$, and the estimation of endmember $E_t$.

## C. Improvements for Dynamical Discretization

In this subsection, we make improvements on Eq.(21) by proposing multistep and bidirectional discretization approaches. These improvements can enable the efficient collection of information from diverse temporal and directional perspectives.

*1) Multistep Discretization Approach:* Latent ODEs model the evolution of latent variables over time in a continuous manner. Consequently, they can naturally accommodate and handle arbitrary time gaps between observations, allowing for the inclusion of irregularly spaced observations. In order to preserve this important property as much as possible after discretization, we propose multistep discretization approach.

Based on eq.(21), we take an arbitrary time interval as K. Calculating a distant time interval directly may sometimes introduce errors, thus we consider $\Delta = 1 : K$:

$$A_t = F_{t-K}(z_{t-K}) + z_t$$
$$\ldots \qquad (22)$$
$$A_t = F_{t-1}(z_{t-1}) + z_t$$

By incorporating all information through a simple yet effective averaging model:

$$A_t = \frac{\sum_{i=1}^{K}(F_{t-i}(z_{t-i})) + K \times z_t}{K} \qquad (23)$$

Multistep discretization is especially valuable to manage arbitrary time gaps between observations. This property making it well-suited for not uniformly spaced hyperspectral sequence images. Moreover, multistep discretization can effectively process and integrate more information from multitemporal signals, allowing it to take into account images across various temporal scales.

*2) Bidirectional Discretization Approach:* Eq.(23) is obtained by following the left-to-right direction, which does not constrain how the previous hidden state abundance communicates with the next image input. In order to establish superior communication across temporals than left-to-right one direction, we consider right-to-left direction simultaneously to construct bidirectional Discretization.

Abundance $A_t^L$ obtained from left-to-right and abundance $A_t^R$ obtained from right-to-left are expressed as:

$$A_t^L = \frac{\sum_{i=1}^{K}(F_{t-i}(z_{t-i})) + K \times z_t}{K}$$
$$A_t^R = \frac{\sum_{i=1}^{K}(F_{t+i}(z_{t+i})) + K \times z_t}{K} \qquad (24)$$

According to both left-to-right direction and right-to-left direction, we design a new bidirectional discretization approach through a simple yet effective additive model:

$$A_t = \frac{A_t^L + A_t^R}{2}$$
$$= \frac{\sum_{i=1}^{K}[(F_{t-i}(z_{t-i})) + (F_{t+i}(z_{t+i}))]}{2K} + z_t \qquad (25)$$

Bidirectional discretization approach allows the model to utilize both preceding and following input signals. This capability leads to a richer and more comprehensive understanding of each abundance. Bidirectional discretization is particularly valuable when the abundance heavily relies on both its preceding and subsequent hyperspectral images. This approach enables MiLD to capture the temporal dependencies across the entire sequence of observations.

## D. Overall Algorithm for Problem Solving

In this subsection, to give a solution of multitemporal unmixing problem, we propose an algorithm which parameterizes the above latent dynamics and discretization operations by a neural network. The whole algorithm and loss functions are introduced together.

To construct an architecture that allows for capturing the evolution of materials, we extend the multistep and bidirectional discretization approach from Section III-C by proposing multitemporal latent dynamics algorithm (see Fig.3). This algorithm consists of three parts from bottom to top: the calculation of pseudo-abundances, the latent dynamical layer for abundances, and the estimation of endmembers. The stability of algorithm will be proved in Section III-D.

Autoencoders are introduced to obtain pseudo-abundances $\approx_t$ and provide latent spaces. We develop T encoders to capture different features from each time respectively, following eq.(10). Consistent with traditional autoencoder unmixing methods, weights of linear decoders are viewed as endmembers based on linear mixing models.

New dynamics of abundances are constructed by developing eq.(25). The resulting complex unit is called latent dynamics layer (see Fig.3). Each layer $\Phi_t(\cdot)$ at time $t$ is built by multistep and bidirectional feature fusion parts with input signal extraction. To describe complex discretization approach, we construct a neural network structure inspired by universal approximation theorem:

*Lemma 1 (universal approximation theorem):* For any function $f$ and a criteria of closeness $\epsilon > 0$, if there are enough neurons in a neural network, then there exists a neural network with many neurons that does approximate $f$ to within $\epsilon > 0$. Then we have sufficient evidence to construct feature extraction networks $F_{1:T}$ at each time $t$, which satisfy $F_t(\approx_t) = \frac{\Delta}{2}\left[2 \times G^2(z_t) + \Delta \frac{d(G^2(z_t))}{d(t)}\right]$. Thus abundance $A_t$ at time $t$ can be obtained by:

$$A_t = L_t(\approx_{t-K}, \ldots, \approx_t, \ldots, \approx_{t+K})$$
$$= \frac{F_-(z_-) + F_+(z_+)}{2K} + z_t \quad (26)$$
$$\doteq \Phi_t(z_t, y_t, K)$$

where $F_{\pm}(\approx_{\pm}) = \sum_{i=1}^{K}(F_{t\pm i}(\approx_{t\pm i}))$. It is noteworthy that if one side of $A_t$ is not contain sufficient K steps, we will complement the remaining steps to reach a total of 2K steps from the other side. For example, at time $t = 1$, we take $A_1 = \frac{\sum_{i=1}^{2K}(F_{1+i}(z_{1+i}))}{2K} + z_1$. At time $t = T - 1$, we take $A_{T-1} = \frac{\sum_{i=1}^{2K-1}(F_{T-1-i}(z_{T-1-i}))}{2K} + \frac{F_T(z_T)}{2K} + z_{T-1}$.

Moreover, since $E_t$ are related intrinsically but varies temporally. We take the traditional PLMM for the estimation of endmembers, which is mentioned in eq.(4). Penalization function is presented to constrain endmembers, leading to:

$$\mathcal{L}_E(E_t) = \frac{1}{2}\|\bar{E} - E\|_F^2 \quad (27)$$

In the design of total loss, we take reconstruction loss and endmember variability contraints to train our model.

$$\mathcal{L}_{RE}(Y_t, A_t, E_t) = \sqrt{\frac{1}{T}\sum_{t=1}^{T}\|Y - A_t E_t\|^2} \quad (28)$$

$$L(Y_t, A_t, E_t) = \alpha L_{RE} + \beta L_E \quad (29)$$

### E. Theoretical Support for MiLD

Consistency, convergence, and stability are essential properties for the proposed MiLD framework, and demonstrating these properties can provide theoretical guarantees. Consequently, in this subsection, we will demonstrate these three properties. Specifically, consistency will show that the discretization strategy we propose does not change the final results. Convergence ensures the correctness of our method for approximating abundance by using latent variables. Stability indicates that even if the initial unmixing results are suboptimal, they will not negatively affect the performance of subsequent unmixing results.

Detailed mathematical definitions and proofs of these properties are given below. Before we proceed further, we shall state some fundamental assumptions.

*Assumption 1 (Continuity of Material Change):* We assume that the material properties of the scene exhibit minimal variation over very short time intervals. The discreteness of material changes is arise from sufficiently large temporal gaps between observations. As $G^2(A_t)$ aims to learn the hidden states, the continuity with a small time interval can be expressed as $\lim_{\Delta \to 0} G^2(A_t) = 0$. The discreteness with a large time interval can be expressed as $|A_t - A_{t-K}| > \Pi$. *Assumption 2 (Lipschitz Condition):* We assume that $G^1(\cdot)$, $G^2(\cdot)$ satisfy the Lipschitz Condition that $|G^1(x) - G^1(y)| \leq L^1|x - y|$ and $|G^2(x) - G^2(y)| \leq L^2|x - y|$, $L^1$ and $L^2$ are constant.

*1) Consistency:* Local truncation error introduced in the stage of computation can produce unstable numerical result. In order to avoid this situation, we need to verify our discretization approach is consistent, that is a very small local truncation error.

*Definition 1 (Consistency):* Let $\rho_t$ denote the local truncation error at the t-th step of the numerical method. The method is said to be consistent with the differential equation it approximates if:

$$\lim_{\Delta \to 0} \max_{1 \leq t \leq T} |\rho_t| = 0 \quad (30)$$

*Theorem 1 (Consistency):* Eq.(19) is said to be consistent with the differential equation eq.(11).

*Proof:* The local truncation errors between eq.(19) and eq.(11) because of the Euler discretization and Taylor expansion in eq.(15), eq.(16) and eq.(18). We denote $G''(\cdot)$ as the second-order term of $G(\cdot)$, thus local truncation errors can be expressed.

In eq.(15), local truncation error is

$$\rho_t^1 = \frac{\Delta^2}{2}G''(A_t, Y_t, t) \quad (31)$$

In eq.(16), local truncation error is

$$\rho_t^2 = \frac{\Delta^2}{2}G^{1''}(Y_t) \quad (32)$$

In eq.(18), local truncation error is

$$\rho_t^3 = \frac{\Delta}{2} \times \frac{\Delta^2}{2} G^{2\prime\prime}(A_t) \qquad (33)$$

Thus

$$\rho_t = \rho_t^1 + \rho_t^2 + \rho_t^3 \\ = \Delta^2 \left[ G''(A_t, Y_t, t) + \frac{2-\Delta}{4} G^{2\prime\prime}(A_t) \right] \qquad (34)$$

is of order $O(\Delta^2)$. Thus $\lim_{\Delta \to 0} \max_{1 \leq t \leq T} |\rho_t| = 0$ is verified.

*2) Convergence:* Intuitively this says that for very small step sizes, the maximum error at any time $t$ between the approximation $\approx_t$ and the true solution $A_t$ is very small.

*Definition 2 (Convergence):* A sequence $\approx_t$ is said to be convergent to another sequence $A_t$ if $\forall \varepsilon > 0$:

$$|A_t - \approx_t| < \varepsilon \qquad (35)$$

*Theorem 2 (Convergence):* $\approx_t$ is convergent to $A_t$ if $\Delta$ is consider very small, that is:

$$\lim_{\Delta \to 0} |A_t - z_t| < \varepsilon \qquad (36)$$

*Proof:* We proceed by mathematical induction to prove $\lim_{\Delta \to 0} |A_t - z_t| = 0$.

First, for the base case $t = 1$, we show that $\lim_{\Delta \to 0} |A_1 - z_1| = 0$.

Next, we assume that for some integer $t - \Delta$, $\lim_{\Delta \to 0} |A_{t-\Delta} - z_{t-\Delta}| = 0$ holds true. That is our inductive hypothesis.

Finally, we demonstrate that $\lim_{\Delta \to 0} |A_t - \approx_t| = 0$ by using the inductive hypothesis and deriving a relationship between $\approx_t$ and $A_t$ as mentioned in eq.(17).

$$\lim_{\Delta \to 0} |A_t - z_t| \\ = \lim_{\Delta \to 0} \left| A_{t-\Delta} - z_{t-\Delta} + \frac{\Delta}{2} [G^2(A_t) + G^2(A_{t-\Delta})] \right| \\ \leq \lim_{\Delta \to 0} |A_{t-\Delta} - z_{t-\Delta}| + \lim_{\Delta \to 0} \frac{\Delta}{2} [G^2(A_t) + G^2(A_{t-\Delta})] \\ = 0 \qquad (37)$$

last equal sign is true because of the inductive hypothesis and Assumption 1. Thus $\lim_{\Delta \to 0} |A_t - \approx_t| = 0 < \varepsilon$ holds.

*3) Stability:* The preceding theorem validates that our derivation process is both reasonable and rigorous. Since stable and unstable equilibria play quite crucial roles in the dynamics of a system, it is useful to be able to classify equilibrium points based on their stability. In the following, we will verify the stability of the proposed latent dynamical system.

*Definition 3 (Stability) [53]:* Let $A_t^1$ and $A_t^2$ denote two different solutions of differential eq.(11) with the initial condition specified as $z_0^1$ and $z_0^2$ respectively, such that $|z_0^1 - z_0^2| < \varepsilon$, $\varepsilon > 0$. If the two numerical estimates are generated by

$$A_t^1 = \Phi(z_t^1, y_t, K) \qquad (38)$$
$$A_t^2 = \Phi(z_t^2, y_t, K) \qquad (39)$$

The condition that for a constant $N$

$$|A_t^1 - A_t^2| \leq N |z_0^1 - z_0^2| \qquad (40)$$

is the necessary and sufficient condition that the latent dynamics system be stable.

*Theorem 3 (Stability):* MiLD is a stable dynamics system.

*Proof:*

$$|A_t^1 - A_t^2| \\ = \frac{\sum_{i=1}^{K} [(F_{t-i}(z_{t-i}^1)) + (F_{t+i}(z_{t+i}^1))]}{2K} + z_t^1 \\ - \frac{\sum_{i=1}^{K} [(F_{t-i}(z_{t-i}^2)) + (F_{t+i}(z_{t+i}^2))]}{2K} + z_t^2 \qquad (41)$$

Because input $Y_t$ is the same at the same time,

$$|z_\Delta^1 - z_\Delta^2| = |z_0^1 - z_0^2| + \Delta G^1(Y_0^1) - \Delta G^1(Y_0^2) \\ = |z_0^1 - z_0^2| = |z_t^1 - z_t^2| = |z_{t-\Delta}^1 - z_{t-\Delta}^2| \qquad (42)$$

We

$$F_t(z_t^1) - F_t(z_t^1) \\ = \frac{\Delta}{2} \left[ 2G^2(z_t^1) - 2G^2(z_t^2) + \Delta \frac{d(G^2(z_t^1))}{d(t)} - \Delta \frac{d(G^2(z_t^2))}{d(t)} \right] \\ = \frac{\Delta}{2} [G^2(z_t^1) + G^2(z_{t-\Delta}^1) - G^2(z_t^2) - G^2(z_{t-\Delta}^2)] \\ \leq \frac{L^2 \Delta}{2} [|z_t^1 - z_t^2| + |z_{t-\Delta}^1 - z_{t-\Delta}^2|] \\ = L^2 \Delta |z_0^1 - z_0^2| \qquad (43)$$

Then we can get

$$|A_t^1 - A_t^2| \leq (L^2 \Delta + 1) |\approx_0^1 - \approx_0^2| \qquad (44)$$

where $L^2\Delta+1$ is a constant. Thus MiLD is a stable dynamics system.

## IV. Experiment

To evaluate the effectiveness of proposed MiLD, we conducted experiments using one real dataset and two synthetic datasets, comparing it against six benchmark algorithms. We considered two evaluation metrics in both qualitative and quantitative manners.

### A. Experiment Settings

We adjust hyperparameters for achieving appropriate endmember and abundance estimations. For the experimental sets under consideration, we set the number of steps $K$ in Eq.(26) to 2 and the number of epochs during the model training to 200. The proposed method is implemented in Pytorch and run in VsCode with a 3090 GPU.

In order to accurately measure the results of the experiment, the quantitative performance is expressed by the value of normalized root mean square error (NRMSE). We take NRMSE of abundance ($NRMSE_A$) and reconstructed image ($NRMSE_Y$) as evaluation metrics.

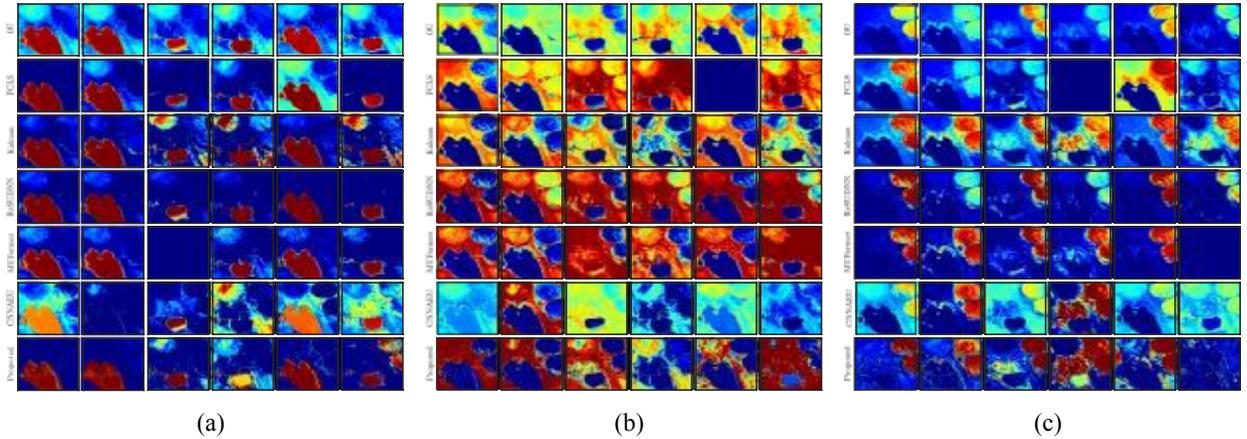

Fig. 4. Comparison of abundance maps on Lake Tahoe dataset. (Top to bottom) Abundance maps obtained by groundtruth, FCLS, OU, Kalman, ReSUDNN, Muformer, CNNAEU and proposed MiLD.The top row represents the true abundances of six phrases. (Left to right) Multitemporal abundance maps with water, soil, and vegetation endmembers.

$$NRMSE_A = \left(\frac{1}{T}\sum_{t=1}^{T}\sum_{n=1}^{N}\frac{\|a_{n,t} - \hat{a}_{n,t}\|^2}{\|a_t\|^2}\right)^{\frac{1}{2}} \quad (45)$$

$$NRMSE_Y = \left(\frac{1}{T}\sum_{t=1}^{T}\sum_{n=1}^{N}\frac{\|y_{n,t} - \hat{a}_t\hat{E}_{n,t}\|^2}{\|y_{n,t}\|^2}\right)^{\frac{1}{2}} \quad (46)$$

where $a_{n,t}$ is the true abundance value of $n$th pixel at $t$ time, $\hat{a}_{n,t}$ represents the estimated abundance. $y_{n,t}$ is the original observed image of $n$th pixel at $t$ time. The reconstructed image is obtained by multiplying $\hat{a}_t$ and $\hat{E}_{n,t}$, where $\hat{E}_{n,t}$ is predicted endmember.

We compare our method with the fully constrained least squares (FCLS), online unmixing (OU) [29], Kalman filter and expectation maximizationbased strategy (referred to simply as Kalman) [31], ReSUDNN based on variational RNN for dynamic unmixing [33], MUFormer based on transformer [35] and an autoencoder-based method CNNAEU for single phase unmixing (referred to simply as single) [37]. When conducting single-phase unmixing algorithm, we regard multitemporal problem as $T$ independent unmixing task and perform CNNAEU at every phrase image. The initialization endmembers of the above methods are obtained by the VCA algorithm.

### B. Experiments on Real Dataset

The Lake Tahoe dataset was acquired by the airborne Visible Infrared Imaging Spectrometer (AVIRIS) during the period from 2014 to 2015 (see Fig.5). The dataset comprises hyperspectral images captured at six distinct temporal phases, with each image possessing a resolution of $150 \times 110$ pixels. After removing the water absorption band, the remaining 173 spectral bands were subjected to further analysis. It is noteworthy that the dataset encompasses three distinct endmembers, namely water, soil, and vegetation, each of which exhibits unique temporal variations.

Due to the unavailability of grounttruth for the Lake Tahoe dataset, we present a qualitative evaluation of the abundance map, with the results illustrated in Fig. 4. Upon inspection, it is evident that most of the comparison methods yield generally satisfactory results, with the abundance distributions appearing reasonable across the majority of the image. However, the proposed method, MiLD, demonstrates superior performance in specific regions, particularly in the upper-left corner of the image, where the abundance of the first and second endmembers is more accurately represented. This improvement highlights the effectiveness of the proposed MiLD for better capturing the variations in endmember abundances over time.

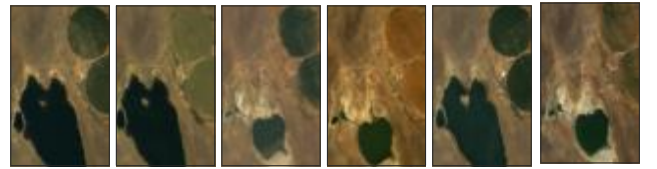

Fig. 5. Lake Tahoe dataset. (Left to right) Moments are obtained from 04/10/2014, 06/02/2014, 09/19/2014, 11/17/2014, 04/29/2015, 10/13/2015, respectively

### C. Experiments on Synthetic Datasets

In order to accurately test the effect of our proposed model, we expand the test on two synthetic datasets, including Synthetic data1 and Synthetic data 2. We adopt both qualitative metrics and quantitative visualizations to assess the results.

*1) Dataset Description:* Synthetic data1 consists of six time hyperspectral synthetic images, each image contains $H \times W = 50 \times 50$ pixels. Endmembers were taken at random from the USGS library as the reference endmember matrixs. The number of endmember is set as $P = 3$ with bands $L = 224$. To model the variability of the endmembers, the reference signatures of each pixel shall be multiplied with a piecewise linear random scaling factor of the amplitude range $[0.85, 1.15]$. Gaussian noise with SNR=30dB was added to simulate realistic scenarios, and local pixel mutations were added to $t \in 2, 3, 4, 5$ to better match endmember changes. This dataset was generated first by literature [33].

Synthetic data 2 contained $T = 15$ sequence images, each of them has $H \times W = 50 \times 50$ pixels and the number of endmembers is set as P = 4. To generate multitemporal imags, a sequence of abundance maps randomly generated according to a Gaussian random field and small, spatially compact abrupt changes was considered. To introduce realistic spectral variability, the signatures at each pixel and time point were selected at random from the set of pure pixels of water, vegetation and roads, which were manually extracted from the Jasper Ridge imagery, using L = 198 bands. The image sequence was then generated with white Gaussian noise at 30-dB SNR according to the multitemporal LMM.

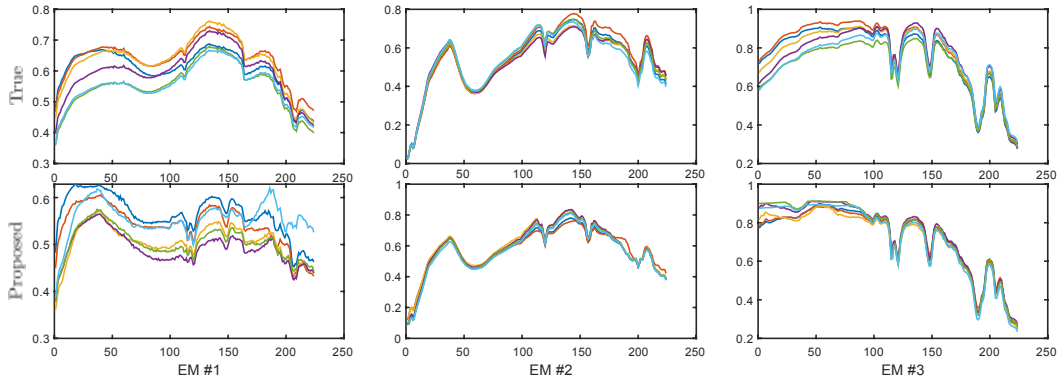

Fig. 6. The endmember estimation results on Synthetic data 1. The top row represents the true endmember results and the bottom row is the result of the proposed MiLD. Different color lines describe multiple phrases. (Left to right) Endmember 1, 2 and 3.

TABLE I
QUALIATIVE RESULTS OF FCLS, OU, KALMAN, RESUDNN, MUFORMER, CNNAEU AND PROPOSED ON SYNTHETIC DATASETS.

| Dataset | Metrics | FCLS | OU | Kalman | ReSUDNN | MUFormer | CNNAEU | proposed |
|---|---|---|---|---|---|---|---|---|
| Synthetic data1 | NRMSE$_Y$ | 0.086 | 0.059 | 0.061 | 0.089 | 0.079 | 0.094 | **0.057** |
|  | NRMSE$_A$ | 0.537 | 0.434 | 0.356 | 0.318 | 0.255 | 0.442 | **0.243** |
| Synthetic data2 | NRMSE$_Y$ | 0.122 | **0.055** | 0.108 | 0.160 | 0.153 | 0.200 | 0.191 |
|  | NRMSE$_A$ | 0.500 | 0.335 | 0.659 | 0.294 | **0.185** | 0.415 | 0.307 |

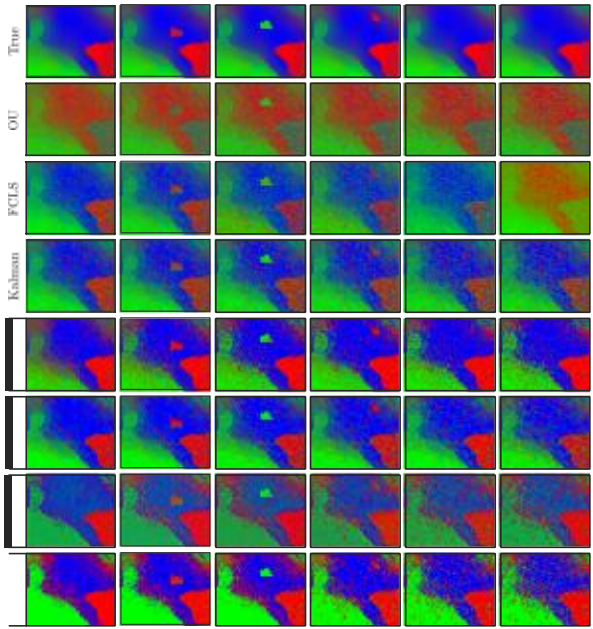

Fig. 7. Comparison of abundance maps on Synthetic data 1. (Top to bottom) Abundance maps obtained by groundtruth, FCLS, OU, Kalman, ReSUDNN, MUFormer, CNNAEU and proposed MiLD. The top row represents the true abundances of six phrases. (Left to right) Abundance maps obtained from moment 1 to moment 6. Green, blue, and red represent the three endmembers, respectively.

*2) Qualitative Result:* We performed a comprehensive com - parative evaluation of our proposed method against six dif- ferent benchmark approaches on the syn1 and syn2 datasets, as detailed in Tab.(I). The performance of each method was assessed using two key evaluation metrics: NRMSE$_A$ and NRMSE$_Y$. These metrics provide a quantitative measure of how accurately the algorithms estimate the abundance maps and reconstruct the hyperspectral images.

Our method delivered the best performance on the first dataset, showcasing the superior capability of our algorithm in both abundance estimation and the reconstructed image qual- ity. Specifically, the abundance maps generated by our method exhibit a high degree of accuracy, while the reconstructed images display more faithful spectral reconstruction compared to the other approaches. For the endmember representation on the first dataset, we will present visual comparisons. While our method outperformed the others on the first dataset, we acknowledge that it did not achieve the best results on the second dataset. However, to provide a more comprehensive evaluation, we will include specific abundance comparison images in the visualization section.

*3) Quantitative Result:* In addition to the quantitative met - rics, we present visual comparisons of the endmember repre- sentations on both synthetic data 1 and synthetic data 2.

On synthetic data 1, we display the results of the abun- dance maps for each model in Fig.7 and the results of the endmembers in synthetic dataset 1 in Fig.6. Among them, Fig.7 shows the comparison of abundance maps estimated by different algorithms. Fig.6 shows the true endmember and the estimated endmember from MiLD. The closer the estimated image is to the real image, the better the performance of the algorithm. Since the synthetic dataset 2 contains 15 phases, we only compare the MiLD estimated abundance maps with the real abundances at the first endmember, taking into account space constraints. The comparison results are shown in Fig.8.

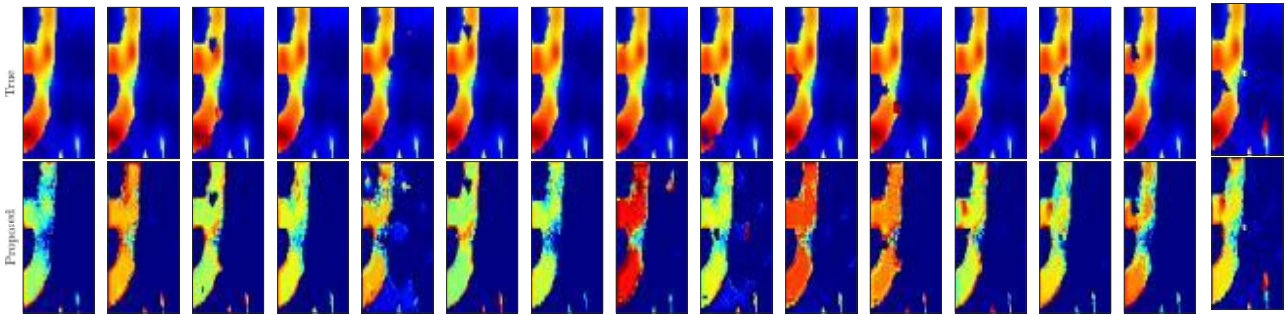

Fig. 8. Estimated abundances of first endmember on Synthetic data 2. The top row represents the true abundance results and the bottom row is the result of the proposed MiLD. (Left to right) Abundance maps obtained from moment 1 to moment 15.

## V. CONCLUSION

In this paper, we propose a multitemporal latent dynamics (MiLD) unmixing framework. MiLD defines, models, solves and supports multitemporal unmixing by capturing the evolution of material dynamics with theoretical analysis. The problem definition is formulated by ordinary differential equations which are conducted by latent variables. The problem modeling is obtained by dynamical discretization approaches which develop mathematical expansions to describe discreteness of images. The problem is solved by proposed algoriothm which captures endmember variability and abundance evolution in latent space. Moreover, the above process are supported by the demonstration of consistency, convergence and stability, all of them are crucial properties for ordinary differential equations. The major contributions of neural MiLD include defining problem by ordinary differential equations, modeling problem by dynamical discretization approach, solving problem by multitemporal unmixing algorithm, and presenting theoretical support. Experiments conducted on both synthetic and real datasets have substantiated the usefulness of the work.